\documentstyle[epsfig,preprint,aps]{revtex}
\advance\hsize by 0.0truecm  \hoffset=0.7cm
\begin{document}
\bibliographystyle{prsty}
\newcommand{\beq}{\begin{equation}}
\newcommand{\eeq}{\end{equation}}
\newcommand{\bea}{\begin{eqnarray}}
\newcommand{\eea}{\end{eqnarray}}
\newcommand{\ms}{m_{\rm s}}
\newcommand{\alf}{{\bar a}}
\newcommand{\bet}{{\bar b}}
\newcommand{\dmuu}{{\partial^\mu}}
\newcommand{\dmud}{{\partial_\mu}}
\newcommand{\lb}{\hfil\break }
\newcommand{\qeq}[1]{eq.\ (\ref{#1})  }
\newcommand{\qeqs}[2]{eqs.\ (\ref{#1}) and (\ref{#2}) }
\newcommand{\queq}[1]{(\ref{#1})}
\newcommand{\qutab}[1]{Tab. (\ref{#1}) }
\newcommand{\qufig}[1]{Fig. (\ref{#1}) }
\newcommand{\qtab}[1]{Tab. (\ref{#1})}
\newcommand{\dsl}{ \rlap{/}{\partial} }
\newcommand{\vsl}{\rlap{V}{\hskip1pt/}}
\newcommand{\asl}{\rlap{A}{\hskip2pt/}}
\newcommand{\zsl}{\rlap{z}{/}} 
\newcommand{\xsl}{\rlap{x}{/}}
\newcommand{\pst}{ \rlap{/}{p} }
\newcommand{\half}{\frac{1}{2}}
\newcommand{\quref}[1]{\cite{ins:#1}}
\newcommand{\qref}[1]{Ref.\ \cite{bolo:#1}}
\newcommand{\wu}{\sqrt{3}}
\newcommand{\nn}{\nonumber \\ }
\newcommand{\Y}{\ {\cal Y}}
\newcommand{\Sp}{{\rm Sp\ } }
\newcommand{\Tr}{{\rm Tr}_{\gamma\lambda_c}}
\newcommand{\Trto}{{\rm Tr}_{\gamma\lambda_c(to)}}
\newcommand{\tr}{{\rm tr} }
\newcommand{\sign}{{\rm sign} }
\newcommand{\Spto}{{\rm Sp_{(to)}\ } }
\newcommand{\linie}{\ \vrule height 14pt depth 7pt \ }
\newcommand{\intT}{\int_{-T/2}^{T/2} }
\newcommand{\Nc}{N_{\rm c}}
\newcommand{\Ne}{$N_{\rm c}$\ }
\newcommand{\gs}{$g_{\rm A}^{0}$}
\newcommand{\gt}{$g_{\rm A}^{3}$}
\newcommand{\go}{$g_{\rm A}^{8}$}
\newcommand{\Gs}{g_{\rm A}^{0}}
\newcommand{\Gt}{g_{\rm A}^{3}}
\newcommand{\Go}{g_{\rm A}^{8}}
\newcommand{\MeV}{{\rm\ MeV}}
\newcommand{\lag}{Lagrangian }
\newcommand{\uas}{$U_A(1)$ symmetry }
\newcommand{\atan}{\rm arctan }
\newcommand{\intq}{\int {d^4q\over {(2\pi)}^4}  }
\newcommand{\daa}{D_{88}^{(8)}(A) }
\newcommand{\dari}{\sum_{i=1}^3 D_{8i}^{(8)}(A) R_i }
\newcommand{\dara}{\sum_{a=4}^7 D_{8a}^{(8)}(A) R_a }
\newcommand{\daiai}{\sum_{a=1}^3 D_{8i}^{(8)}(A) D_{8i}^{(8)}(A) }
\newcommand{\daaaa}{\sum_{a=4}^7 D_{8a}^{(8)}(A) D_{8a}^{(8)}(A) }
\newcommand{\bit}{\begin{itemize}}
\newcommand{\eit}{\end{itemize}}
\newcommand{\tot}{ {3\over 2} }
\newcommand{\ba}{\begin{array} }
\newcommand{\ea}{\end{array} }
\newcommand{\ksl}{ \rlap{/}{k} }
\newcommand{\ddmu}{\partial_\mu }
\newcommand{\dumu}{\partial^\mu }
\newcommand{\vx}{ {\vec x}  }
\newcommand{\seff}{S_{\rm eff} }
\newcommand{\phans}{\phantom{abc} }
\newcommand{\phan}{\phantom{abcdefgh} }
\newcommand{\A}{ {\cal R} }
\newcommand{\njlm}{Nambu--Jona-Lasinio model\ }
\newcommand{\njl}{Nambu--Jona-Lasinio\ }
\newcommand{\oon}{$1/N_c$}
\newcommand{\GeV}{{\rm\ GeV}}
\newcommand{\disp}{\displaystyle }
\newcommand{\vy}{ {\vec y} }
\newcommand{\vz}{ {\vec z} }
\newcommand{\dtau}{\partial_\tau}
\newcommand{\Mhat}{{\hat M}}
\newcommand{\cor}{{\cal C}_B(T)}
\newcommand{\brax}{\langle \vx,t_x\mid}
\newcommand{\bracx}{\langle x\mid}
\newcommand{\ketsx}{\mid x\rangle}
\newcommand{\bray}{\langle \vy,t_y\mid}
\newcommand{\braz}{\langle \vz,t_z\mid}
\newcommand{\ketx}{\mid\vx,t_x\rangle}
\newcommand{\kety}{\mid\vy,t_y\rangle}
\newcommand{\ketz}{\mid\vz,t_z\rangle}
\newcommand{\defined}{\stackrel{\rm def}{=}}
\newcommand{\signum}{{\rm sign} }
\newcommand{\ket}[1]{\mid{#1}\rangle}
\newcommand{\bra}[1]{\langle{#1}\mid}
\newcommand{\mata}{\left( \ba{c} }
\newcommand{\mate}{\ea \right) }
\newcommand{\diag}{{\rm diag}}
\newcommand{\sumi}{ \sum_{i=1}^3 }
\newcommand{\suma}{ \sum_{a=4}^7 }
\renewcommand{\arraystretch}{1.5}
\newcommand{\fm}{{\rm\  fm}}
\newcommand{\ohatz}{{\hat O}_\mu(z)}
\newcommand{\square}{\partial_i^2}
\newcommand{\PP}{ {\rm PP} }
\newcommand{\moment}{ \Theta_U }
\newcommand{\bmu}{{\bar\mu}}
\newcommand{\bmo}{{\bar m}_0}
\preprint{SUNY-NTG-9-96}
\title{Pion electromagnetic formfactor   \\
        in the instanton vacuum }
\author{    Andree Blotz
         \footnote{email:andree.blotz@sunysb.edu }
           and Edward Shuryak
         \footnote{email:edward.shuryak@sunysb.edu }
         }
\address{
Department of Physics, State University of New York in Stony Brook, \\
NY  11794-3800, USA
      }
\date{ \today  \\
SUNY-NTG-9-96
  }
\maketitle
\begin{abstract}

We have calculated the 3-point
correlation function for the electromagnetic interaction of the pion
in an ensembles of instantons and
anti-instantons, modelling
the QCD vacuum. The results are well described by a pion pole and
the pion formfactor is extracted, nicely following a
standard
monopole fit. 
The experimental data on the formfactor 
are well reproduced, provided  an average instanton size is fixed at  
${\bar\rho}=0.35\pm 0.03\fm$, the same as found
for a variety of 
other correlation functions and was found on the lattice.

\end{abstract}
\section{Introduction}

   Significant progress has been made during the last few years 
in understanding of the  QCD vacuum and hadronic structure
in terms of instantons. 
In a series of papers 
\cite{ins:es93a,ins:es93b,SSV_94} it  has been shown
that a specific model, 
which assumes ensembles of instantons
and anti-instantons in the QCD vacuum,  can not only describe
the gross features of
the vacuum, such as spontaneous chiral symmetry breaking or 
topological 
susceptibilities\footnote{See e.g. \cite{ins:divar} 
for a recent review
of chiral symmetry breaking 
mechanism by instantons}, but can also $quantitatively$ 
describe a large
number of mesonic and baryonic correlation functions, 
 in agreement with phenomenlogy \cite{ins:phen}
and lattice calculations \cite{ins:negele}. Furthermore, the parameters of the
ensemble have been confirmed by lattice studies \cite{CGHN_94,MS_95a}, 
and also
the dominant role of instantons was directly demonstrated by removal of
all other kinds of the gauge fields from the configurations by the so called
{\it cooling} procedure \cite{CGHN_94}.

  In view of those developments, wider practical applications of
 the instanton-based models are justified. Since studies of 2-point
correlators have produced hadronic masses
 and coupling constants in agreement with data,
it is natural to perform more 
detailed studies and test
whether those models  can or cannot describe details of hadronic
structure as well.     
Three point correlation functions can describe hadronic
 couplings  to external fields, and for this aim they
have been considered  within the QCD sum rule approach
(e.g. \cite{ins:iosm}), the light-cone formalism 
(e.g. \cite{ins:rady,ins:lebr} ),
the generalized impulse approximation within non-perturbative
Dyson-Schwinger equations \cite{ins:craig} and of course on the
lattice \cite{ins:lattice3point,ins:latticepion,ins:lattice3point2}.

Only very recently the first (and the simplest) 3-point
function related with the pion formfactor has been studied in
the so called single-instanton
approach by Forkel and Nielsen \cite{ins:foni}.
For $Q^2\sim 1\GeV^2$ they have found good agreement with data.
Significant advantage of their approach
over the previous QCD sum rule calculation is that
the $pseudoscalar$ (rather than the axial) currents 
are used in order to create
pions.
It makes a significant practical difference, because the pion is
coupled to the former current with large constant $\lambda_\pi$ and to
the latter with the small coupling  $f_\pi$.

Technically Forkel and Nielsen follow closely ref.\cite{Shu_83} in which
the 2-point pion correlator was discussed in the single-instanton
approximation: in it
the propagators are calculated 
using the zero-mode solution of the
Dirac operator in the presence
of one close instanton (or anti-instanton),   
 whereas the effect of the
remote instantons is taken into account in some mean field approximation.
This is a good approximation, provided all
relevant distances $(x,y)$ 
are short
compared to inter-instanton separation 
$R\sim 1\fm >> (x,y)\sim 0.2\fm $. In practice, it is a very
important limitation, since it does not allow to  separate
the pole term from the non-pole or ``continuum" contributions
unambigously. 
 Therefore
 Forkel and Nielsen had to extract the formfactor by a complicated
fitting procedure, similar to what is done in the context of QCD sum rules.
Also, they have performed a traditional Borel transform of the correlator, 
while in fact the whole analysis can better be made directly in
(Euclidean)
space-time representaion, in which all relevant formulae 
are much simpler.

In the present paper we also study with better accuracy
 a variety of
two point correlation functions, and check how the obtained masses
of  hadrons are correlated with hadronic formfactors. The
 electromagnetic form factor of the pion
is of special interest, because the pion is the Goldstone
boson of the spontaneously broken chiral symmetry and therefore
the quark attraction in this channel is very large. 
Furthermore in the models used the pion is bound by instanton-induced forces,
and in a way its formfactor is basically the instanton formfactor.

 We are using the methods developed in \cite{ins:es93a}
which allow to evaluate quark propagators in a multi-instanton
background numerically, up to rather large distances.
Therefore, we are $not$ restricted to small  $x,y << R$ and
thus have no significant ``background" from non-pion states.
As a result, we can extract the pion formfactor with much better accuracy,
and even study how it depends on various modifications of the underlying
ensemble. Since the pion formfactor is also relatively accurately
measured,
those studies proved to provide a very rigid constrained on such
microscopic parameters such as the mean instanton size.

 We  use  two formulations of the statistical instanton-based models:
the simplest    
Random Instanton Liquid Model
(RILM) \cite{ins:es82i,ins:es93a,ins:es93b}
and the much more developed Interacting Instanton Liquid Model (IILM)
\cite{ins:tses95a,ins:tses95b}. While the former one has simply
the previously determined total instanton density and 
average size as input and
assumes a random
distribution in all collective coordinates, 
the latter one is a statistical model which predicts 
correlations between the instantons in the liquid. Some of those
correlations are known to be qualitatively important:
e.g.  quark-induced interactions 
lead  to a screening of the topological 
charge and correct the large distance behaviour of 
the correlators in channels, where the random model 
produces  repulsion \cite{ins:stream}. 
It is also important that in the IILM
the instanton parameters were not fixed but 
determined   
selfconsistently, and after the instanton interaction is fixed the only 
parameter left is $\Lambda_{QCD} $. If its value is fixed from    
an average density of ${\bar n}=1\fm^{-4}$, 
an average instanton size which emerges is somewhat larger
than the original $\rho=1/3\fm$
\cite{ins:es82i} or recent lattice results \cite{CGHN_94,MS_95a}.

The organization of the paper is as follows. In section 
II.  we summarize the phenomenlogical knowledge of the 
3 point correlation function for the pion. In section III. 
we show how to derive this correlator within the Instanton 
Liquid model and show also the various short distance 
approximations to the full simulation. Sect. IV discusses 
the numerical procedure and results for the RILM and  sect. V 
for the IILM. In sect. VI 
we make  
some concluding remarks. The App. A. gives some formulas 
for the direct instanton approach and App. B gives  
formulas for the continuum contribution of the 3 point 
function.

\section{Electromagnetic Pion Form Factor - Notations and Phenomenology  }

The electromagnetic structure of the pion can be measured
in the time-like region, the $\rho$-meson pole containing part
of the formfactor or structure function  and
in the space-like part. In both
cases the formfactor depend
only on the four-momentum transfer of the pion which we denote by
$Q^2=-q^2$. (Below we  always use space-like momenta,
$Q^2 >0$.) 
While the time-like region can be investigated in
collider experiments such as $e^+e^-\rightarrow \pi^+\pi^-$,
the space-like region is obtained from scattering
of pions off atomic electrons (for small $q^2$) or
via the production of pions in $e^-p\rightarrow e^-\pi^+n$
(for larger $q^2$).

 From the theoretical point of view
the interaction of the hadrons with external currents
like the electromagetic one is most
conveniently written in terms
of three point correlation functions. The three points
denote the creation and annihilation points of the hadrons
as well as the location of their interaction with the external current
$j_\mu^Q(y)$:
\beq     
          \Pi_\mu({\bar p},q) =
         \int d^4x \int d^4y\  e^{i {\bar p}  x} \  e^{  i q y }
         \langle 0 \mid  {\cal T} j_5^{\pi\dagger}
         (-x/2)  j_\mu^Q (y) j_5^\pi (x/2)
                   \mid 0  \rangle
\label{g1}
\eeq
where ${\bar p}=(p + p^\prime)/2$ is the average of the momenta
$p,p'$ connected to the pseudoscalar currents.
The charged pions are  represented by the quark bilinears
\beq       j_5^{\pi\pm} (x)  =
           \left(   \begin{array}{c}  {\bar d} (x)i\gamma_5 u(x) \\
            {\bar u} (x)i\gamma_5 d(x)
               \end{array}    \right)
           =   
            {1 \over \sqrt{2} }
            {\bar q} (x) i \gamma_5  \tau^\pm  q (x)
            :=
                {\bar q} (x)  \Gamma_\phi    q (x)
\label{g1a}
\eeq
and
$j_\mu^Q(y) =  {\bar q} (x) Q \gamma_\mu q (x) =
\sum_{i=u,d} e_i {\bar q_i}(y)\gamma_\mu q_i(y)$
is the electromagnetic current.

The 3-point function in the r.h.s. of \qeq{g1} is the quantity we will
calculate below. Its transform, the l.h.s. of \qeq{g1}, 
can be related to physical observables by 
a double dispersion relation \quref{iosm3}\footnote{The
double integral is needed because one has to
use a complete set of physical states both
in the incoming and the outgoing
channel. For a recent discussion on the dispersion relations  cf.
\cite{ins:ioffe3,ins:iosm2}.
Although the \qeq{conttrib} is not well defined due to the
divergent behaviour of the dispersion integral for large
momenta $s,s'$, one can improve that by subtraction of
polynomials
functions of $p^2,p^{\prime 2},Q^2$. } 
\beq
        \Pi_\mu \left( {\bar p},q 
             \right)
          =
        \int_0^\infty  ds_1  \int_0^\infty  ds_2
        {\rho_\mu (s_1,s_2,Q^2=-q^2)
                  \over (s_1-p^2) (s_2 -p^{\prime 2})  }
\label{conttrib}
\eeq
over the spectral density   $\rho_ \mu (s_1,s_2,Q^2=-q^2)=
\rho_1(s_1,s_2,Q^2){\bar p}_\mu +\rho_2(s_1,s_2,Q^2) q_\mu$.
As explained below, the e.m. formfactor of the pion 
can be extracted from the 
former structure containing ${\bar p}_\mu$ and 
$\rho_1(s_1,s_2,Q^2)$.

  A Borel transformation in $p$ and $p^\prime$ was used in earlier
papers, with the  purpose to 
eliminate the polynomial subtraction terms.
Our philosophy is however different: we  make an inverse
Fourier
transfer to coordinate representation instead, in which 
 the propagators
in the instanton background 
are naturally obtained. As shown in details in \cite{ins:phen}
 for the 2-point functions,
the coordinate space is as good as Borel representation for all our 
purposes.\footnote{
As we show in App. B., the
coordinate space representation of \qeq{conttrib} is not only finite
but reproduces also the correct short distant behaviour.}
The second  reason for using the Borel transform in QCD sum rules,
namely to suppress the larger mass resonances, is also not needed here,
because in the coordinate-space correlator
these states are automatically suppressed at distances we will work with.

The spectral density $\rho_\mu(s_1,s_2,q^2)$
is proportional to the
imaginary part of the polarization operator. 
We approximate it as the  pion contribution  given by
\beq  
             \rho_{pole} (s_1,s_2,q^2)  = 
       (2\pi)^3   \delta(s_1 -m_\pi^2)  \delta(s_2 -m_\pi^2)
              \langle 0 \mid j_5(0) \mid \pi \rangle
            \langle \pi \mid j_\mu^Q \mid \pi \rangle
            \langle 0 \mid j_5(0) \mid \pi \rangle
 \label{g10}
\eeq 
Here the overlap matrix element  
\beq     \langle 0\mid j_5(x)\mid\pi(p)\rangle = \lambda_\pi
         (2\pi)^{-3/2} e^{(ipx)} 
\eeq 
is given in terms of 
$\lambda_\pi$, which may be represented as \cite{ins:tses95a}  
\beq   \lambda_\pi = {  {\bar q} q \over \sqrt{2}  f_\pi }
          =
         { \sqrt{2}  {\bar u} u \over  f_\pi }
\label{lamcon}
\eeq
with $f_\pi$ corresponding to the experimental $f_\pi=93\MeV$.
The definition of the electromagnetic form factor $F_\pi(q^2)$ 
\qeq{g10}
is
\beq  
     \langle\pi(p)\mid j_\mu(0) \mid\pi(p')\rangle= e_\pi F_\pi(q^2) 2
{\bar p}_\mu,
\eeq 
which is normalized by $F_\pi(0)=1$. 
Then 
the \qeq{g10} can be written as
\beq       \Pi_\mu({\bar p},q)  =
          { \lambda_\pi^2   2  e_\pi   F_\pi(q^2)
             \over  \left( ({\bar p} -q/2)^2  + m_\pi^2 \right)
                    \left( ({\bar p} +q/2)^2  + m_\pi^2 \right)      }
                             {\bar p}_\mu
         \ \    + \ \  \Pi_{\mu,cont} ({\bar p},q)
\label{poleff}
\eeq
with  free
quark continuum $\Pi_{\mu,cont}({\bar p},q)$ 
(cf. App. B), starting at some threshold value $s_0$.

The experimental measured form factor in the space like region
has been shown to be 
well parametrized by a
monopole form factor \cite{ins:bobek,ins:amen},
resembling is some way the
exchange of a $\rho$ meson in {\it vector dominance} models. 
The parameter in this fit with 
$m_V=679\pm 19\MeV$ \quref{bobek} is   
however 
below the $\rho$ meson mass\footnote{There is also a precise measurement
of the formfactor for smaller $Q^2<0.30\GeV^2$ \cite{ins:amen}, which 
yields, also from a monopole fit, $m_V=736\pm 9\MeV$ and is still below the 
$\rho$ mass.} 
The measurements are performed over the large
range 
\footnote{See also the
CEBAF proposal \quref{CEBAF} for a future
high statistics measurement of the formfactor via
electroproduction
at $Q^2\simeq 0.5-5 \GeV^2$. }  
of space-like $Q^2=0.18-9.77\GeV^2$.
In coordinate space this measures the region around 
$0.05-0.5\fm$. 
The upper limit of which is clearly
dominated by non-perturbative 
effects in the QCD vacuum, as 
will be shown, and in general cannot be
described by the operator product expansion.
The fourier transform of \qeq{poleff}  is then conveniently calculated
from
\bea    \Pi_{\mu,pole} (x,y)
         &=&
        -i \int_0^\infty dq \int_0^1 d\alpha
       { q^2 \over 32 \pi^4 } { J_1 ( q \mid y-x(0.5-\alpha)\mid )
       \over \mid y - x(0.5 - \alpha) \mid   }
       \sqrt{m_\pi^2 +q^2\alpha(1-\alpha)}     \nn & &
       K_1
       \left( \mid x \mid \sqrt{m_\pi^2 +q^2\alpha(1-\alpha) } \right)
       {  2 \lambda_\pi^2  e_\pi   \over 1 + q^2/m_V^2 }
         {x_\mu \over \mid x \mid }
\label{poleff2}
\eea
where $J_1,K_1$ are the normal and modified bessel functions.
The rather cumbersome formula for the continuum is given in
App. B.  In \qeq{poleff2} we used already the
monopole parametrization of the form factor \cite{ins:bobek},
which was shown to be a good description in the 
space like region of the form factor.

\section{The Instanton Liquid }

 The 3-point function in question can be evaluated 
using the same quark propagators in the
 multi-instanton
background fields 
as used for the 2-point functions in
 \cite{ins:tses95a,ins:tses95b}, and we will not
repeat any details here. The generating functional of the IILM can be
written as
\bea   Z[\eta,{\bar\eta},s_\mu^a]
          &=&  { 1 \over N_+!  N_-! }
          \int  \prod_i^{N_+ +N_-}  d \Omega_i d(\rho_i)
          e^{ \disp  - \int d^4 x  d^4 y \ \   {\bar\eta}(x)
          \left[     ({\hat D}
            + m_f -  s_\mu^a \Gamma_\mu^a)(x,y) \right]^{-1}
          \eta(y)   }
           \nn & &
           e^{(-S_{int})} \
          \prod_f^{N_f} \det \left(
         {\hat D}   + m_f -  s_\mu^a \Gamma_\mu^a \right)
\label{general}
\eea    
We will only shortly explain the meaning of the various expressions.
The  ${\hat D}$ is the fermionic Dirac operator in the presence of
instantons and anti-instantons. These instanton solutions are
described by a set of collective coordinates $\Omega_i$, which
are the color orientation $U_i$, the position $z_i$ and the size
$\rho_i$ of the instanton.
Therefore the original gauge measure ${\cal D}A_\mu$ becomes essentially
an integral over the collective coordinates
$d \Omega_i$. The $d(\rho_i)$ in \qeq{general} is the semiclassical 
instanton amplitude, which was originally calculated by 't Hooft: its  
two-loop version is used in \cite{ins:tses95a}.

The $m_f$ are the explicit (chiral and vector)
symmetry breaking current masses and
$S_{int}$  denotes the classical interaction of the instantons, which is
as usual \cite{ins:tses95a} 
approximated by a two-body interaction.

In the case of the electromagnetic formfactor 
one has 
$\Gamma_\mu^Q=\gamma_\mu Q^a \tau^a$, 
where $Q$ is the
electromagnetic charge matrix.
The spin-isospin matrices $\Gamma_\phi$, ($\phi=\pi,\delta$), in \qeq{g1a}
correspond to $i\gamma_5\tau^\pm/\sqrt{2}$
for the charged pions $\pi^\pm$
and to $\tau^\pm/\sqrt{2}$ for the 
isovector scalar $\delta$-meson. 
The reason to
discuss also the $\delta$ shortly will become clear in the following.
From the generating functional \qeq{general} a three point correlator
for a mesonic state is
obtained from
\beq       \Pi_\mu(x,y)      =
            ({\Gamma_\phi})_{ab}^{\alpha\beta}
             ({\Gamma_\phi})_{cd}^{\gamma\delta}
%
           { \delta \over \delta  s^Q_\mu(y) } 
           {  \delta^4   Z[\eta,{\bar\eta},s_\mu^a] \over
            \delta  {\bar\eta}_a^\alpha(-x/2) \delta \eta_b^\beta(-x/2)
            \
           \delta  {\bar\eta}_c^\gamma(x/2) \delta \eta_d^\delta(x/2)
                    }
\label{forms}
\eeq
where latin indices correspond to spin indices and
greek symbols to isospin indices. Since all currents of
interests are color singlets, color indices are implicitly
understood.
In terms of the quark propagators
the general form for the 3-point correlator, using isospin symmetry for
the up and down quark propagator $S_u=S_d$, follows as
\bea  \Pi_\mu(x,y)
       &=&  \left(e_u-e_d \right) \left(    
          \langle
       \tr S(-x/2,y) \gamma_\mu S(y,x/2) \gamma_5 S(x/2,-x/2)\gamma_5
         \rangle
       -      \right. 
       \nn &   &  
       \langle 
       \tr S(y,y)\gamma_\mu \
        \tr S (-x/2,x/2) \gamma_5 S (x/2,-x/2) \gamma_5
           \rangle 
       -       \nn   &  &   \left.  
        \langle 
        \tr S(y,y)\gamma_\mu
         \tr S(-x/2,-x/2) \gamma_5 \ \  \tr S(x/2,x/2) \gamma_5
              \rangle           \right) 
\label{g21}
\eea
Since the quark propagators in \qeq{g21} are evaluated 
in a given instanton anti-instanton background, averaging over
different instanton ensembles        
is implicitly understood by the brackets in \qeq{g21}. 
Here the first term on the RHS of \qeq{g21} is the so called
connected contribution and
the second and third are the disconnected diagrams.
One should note of course that 
the quark propagators are the full propagators
in the background of the multi instanton/anti-instanton
configurations so that the expression 'disconnected' refers only
to the 'quark lines'.
One may note that the disconnected terms are in general 
obtained from the
fermion determinant of the generating
functional \qeq{general} \footnote{ In the case of
effective quark or semi-bosonized quark theories like the NJL
model \cite{ins:review}
these expressions are known as the UV divergent one loop contributions, 
representing the polarization of the vacuum.} .

In the present approach there are two things to mention.
In the coordinate representation
and using the free, massless quark correlator\footnote{Note that we are
using anti-hermitian $\gamma$-matrices, so that \qeq{pfree} is indeed real 
in Euclidean space.} 
\beq   S_0(-{x\over 2},{x\over 2}) = \langle 0 \mid {\cal T}
       u(-{x\over 2}) {\bar u} ({x\over 2}) \mid 0 \rangle
                =   
       -   { i \over 2 \pi^2 } { \xsl \over x^4 }
\label{pfree}
\eeq
this second and third terms in \qeq{g21} 
would have a short-distance singularity. In the case of the 
vector current, one has to subtract the 
free propagator \cite{ins:gesh}.\footnote{ 
In the presence of the instanton also the non-zero 
modes get distorted \cite{BCC_78,ins:es93a} and 
some disconnected contributions 
still survive and are even divergent \cite{ins:es93a}. They however
disapper
after insertion of  a path-ordered exponential 
$          {\bar q} (y) Q \gamma_\mu
           P\exp{(-i\int_y^{y+\epsilon} A_\mu(z) dz_\mu) }
          q(y+\epsilon)
$ 
which makes the
electromagnetic current gauge invariant and conserved.}
The disconnected parts can be shown to vanish (which follows already from
current conservation \quref{gesh}) and only the
triangular diagram of \qeq{g21} survives.

For very $small$ Euclidean space-time
separations
$\mid x\mid,\mid y\mid\ll 0.2\fm$
the 3-point function
$\Pi_\mu(x,y)$
is governed by the free
Dirac propagator \qeq{pfree}.
For $x\cdot y=0$, which we use, it is especially simple
\beq        \lim_{x,y \to 0}     
            \Pi_\mu (x,y)        
             =  -
        {N_c e_\pi \over 2 \pi^6 } 
        {i x_\mu\over x^4 } {1\over (y^2 + x^2/4)^3  }
\label{freeprop}
\eeq 
For larger distances the non-perturbative effects
come in. In the leading order for small 
distances the {\it vacuum dominance} approximation 
would suggest 
the following correction to \qeq{pfree}  
\beq   S(-{x\over 2},{x\over 2}) = - { \langle {\bar u} u\rangle \over
         12 }
\label{vacdom} 
\eeq
It is clear that the term of the order 
${\cal O}\left(\mid{\bar u}u\mid\right)$
vanishes due to spin traces, while
the ${\cal O}(\mid{\bar u} u\mid^2)$ contribution in this approach 
reads\footnote{Note that it is different from the ${\cal O}(\alpha_s
{\langle{\bar\Psi}\Psi\rangle}^2)$ contribution in standard OPE 
analysis \quref{iosm}.}    
\beq       
       \Pi_\mu (x,y) = -    
       {  N_c e_\pi \over 72  \pi^2 }  
        \mid\langle {\bar u} u\rangle  \mid^2  
          { i  x_\mu }  \left( 
          {1 \over (y^2 +x^2/4)^2}   +  { 1 \over x^4 }
              \right) 
\label{nlo}
\eeq              
Compared to \qeq{freeprop}, this next to leading order term 
has the same sign and indicates therefore an enhanced signal 
in this channel. This is similiar to the pion two-point 
correlator itself \cite{ins:es93b} and can be traced back 
to the attractive forces inside  the pion.

However such  "vacuum dominance" approximation 
works only qualitatively and for small distances only.
To go  beyond those  one needs 
a model for the non-perturbative effects.
The basic feature of the instanton-based models  is 
related to the zero mode solutions of the Euclidean 
Dirac operator ${\hat D}$, which are obtained from  
\beq     {\hat D}  \phi_\lambda  
           =  
          \lambda \phi_\lambda, 
\eeq  
for $\lambda=0$  and  exist, if the topology of the gauge 
field configurations is non-trivial. This is the case for 
the instanton solutions.
From these solutions 
we approximate the full Dirac propagator
$S(-x/2,x/2)$ in the
instanton liquid model by a sum of the zero mode contributions
\cite{ins:es93a,ins:es93b}
$S_{ZM}(-x/2,x/2)$:  
\beq      S_{ZM}( -{x\over 2},{x\over 2} )
          =   \sum_{I,J}  \phi_I( -{x\over 2} )
              \langle I  \mid {   1   \over 
               {\hat D}   + m     }  \mid J \rangle  
              \phi_J^\dagger ( {x\over 2} )
\label{zeromodes} 
\eeq
and the non-zero mode term $S_{NZM}(-x/2,x/2)$
\cite{BCC_78,ins:din} 
\beq     S_{NZM}(-{x\over 2},{x\over 2})     
            =  S_0 (-{x\over 2},{x\over 2} )  +
         \sum_{I}
         \left[   
                 S_I(-{x\over 2},{x\over 2})  
           -  
                 S_0(-{x\over 2},{x\over 2} ) 
          \right]
\label{nonzero}
\eeq
where $S_I(-x/2,x/2)$ is the non-zero mode contribution
in the presence of a single 
instanton\footnote{This expression \qeq{nonzero}
should be understood as the leading term in a multiple 
scattering expansion, cf. \cite{ins:es93b} for details.}.

Before discussing the result of the full calculation with 
\qeq{zeromodes} and \qeq{nonzero} in \qeq{g21} 
in an ensemble of instantons, it is instructive to investigate the 
effect of a $single$ instanton, as done by Forkel and Nielsen. 
Then the general form of the 
zero mode propagator is given by   
\beq   
            S_{ZM} ( -{x\over 2},{x\over 2} ) 
            = \sum_\lambda 
          { \Psi_\lambda(-{x\over 2}) 
           \Psi_\lambda^\dagger({x\over 2}) \over \lambda + i m } 
\label{genzero} 
\eeq
where $\Psi_\lambda(x)=\sum_{I=1}^{N}C_I^\lambda\Psi_0^I(x-z_I)$ 
can be expanded in terms of the zero modes  
$\Psi_0^I(x)$ of individual instantons. Now one can 
argue\cite{ins:report} 
that
for small distances the instanton $I_*$ closest 
to ${x\over 2}$ and $-{x\over 2}$ 
dominates so that the propagator can be expressed in terms 
of this single zero mode for an instanton/anti-instanton as    
\bea   
         S_{ZM}^{\rm MF} ( -{x\over 2},{x\over 2} )
          &=&  
                \sum_\lambda   { \mid   C_{I_*}^\lambda \mid^2 \over 
              \lambda + i m  }  
         {   \Psi_0^{I_*} (-{x\over 2}) 
           \Psi_0^{I_*\dagger}
            ({x\over 2})     } 
            \nn &=& 
             { 
          (-{\xsl\over 2} - \zsl) \gamma_\mu \gamma_\nu 
        (  {\xsl\over 2}    -   \zsl)         
             \over  8   {\bar m}  } 
           \left(   \ba{c}  \tau_\mu^- \tau_\mu^+   P_L   \\ 
                        \tau_\mu^+ \tau_\mu^-   P_R    
               \ea    
            \right)    \phi(-{x\over 2}-z) 
                       \phi({x\over 2}-z)               
\label{zeroeff} 
\eea     
where a dynamically generated effective mass ${\bar m}$ 
can be defined from \qeq{genzero} by  
\beq   
            {1 \over {\bar m}  }    = 
              \sum_\lambda   { \mid   C_{I_*}^\lambda \mid^2 \over 
              \lambda + i m  } 
\label{mbardef}  
\eeq       
and 
\beq   
        \phi(x) =   { \rho \over  \pi} {1 \over \mid x\mid 
        \left(  x^2 + \rho^2  \right)^{3/2} } . 
\eeq  
In this approach 
the effect of distant instantons and chiral symmetry breaking is
therefore effectively included by
replacement of
the current quark mass $m$ 
by ${\bar m}$,    
the dynamically generated effective mass \cite{Shu_83}, representing
chiral symmetry breaking by other instantons.    
This was done by Forkel and Nielsen \cite{ins:foni} 
who has used       
the correlator $\Pi_\mu (x,y)$ with \qeq{zeroeff}. 
In the short distance limit\footnote{See our discussion of 
the full expression in the next section and App.A.}
this reduces to 
\beq 
      \lim_{x,y\to 0}
          \Pi_\mu (x,y) = -
       { i e_\pi \over 5 \pi^4 } 
       { {\bar n}  \over {\bar m}^2  \rho^4 }
                    { x_\mu \over (y^2 + x^2/4)^2  }
\label{effshort} 
\eeq
%
%
Note that using  
the mean field results \cite{ins:es82i,ins:dipe}  
for the constituent quark 
mass \footnote{Note that it is not
equal to the so called constituent quark mass 
(and it is in fact about twice
smaller)
because the effective mass in question is not evaluated at zero
momentum. }  
and the condensate, namely       
$\langle{\bar u}u\rangle\sim \sqrt{\bar n}/\rho$ 
and ${\bar m}\sim\sqrt{\bar n}\rho$,  
the eq. (\ref{effshort}) 
gives a correlator which is parametrically larger than 
eq. (\ref{nlo}) by the inverse packing fraction 
$1/f\sim{1}/({\bar n}\rho^4)\gg1$. 
       
%
However for the full Instanton Liquid simulation the formulas of Forkel and 
Nielsen are consistent to our approach for short distances.
But neither the vacuum dominance nor the 
single-instanton approach \cite{ins:foni} can describe
the  correlator for larger distances $x,y\gg 0.2\fm$ necessary for
a clear separation of the pion contribution. 
This goal will be reached in the next section, by using the 
numerically calculated correlators.

\section{Correlators in random instanton vacuum} 

Numerically the propagators are calculated as
solutions to eqs. (\ref{zeromodes}),
(\ref{nonzero}).  Furthermore for the 
RILM we took 256 randomly placed and oriented instantons (half of them
 anti-instantons) 
into a periodic box of $(5.67)^2\times(2.82)^2 \fm^4$, where 
the long box lengths  correspond to the $x,y$ directions, in which the 
correlator is actually measured. 
The numbers are chosen to reproduce an instanton density 
of ${\bar n}=1\fm^{-4}$.  Furthermore,
we take a $variable$ average instanton size   
( $\rho=0.28\fm,0.35\fm,0.42\fm$) in order to show 
the sensitivity of the form factor to the instanton size.
We have evaluated 2- and 3-point functions and studied them in detail.

In Fig.1 we show the two point correlation function.  
According to our pole plus resonance Ansatz (eqs. {\ref{b5},\ref{b6}) 
for the 
correlator we give in the figure the best fit for 
the pion mass $m_\pi$ and the coupling constant $\lambda_\pi$. 
Because of our finite box we used a current quark mass 
of $m_u=m_d=20\MeV$ which is somewhat larger than the 
physical quark masses. Using the Gell-Mann Oakes 
Renner relation, the measured pion mass values 
in Fig. 1 are scaled down to the physical pion mass, 
 see \qutab{tab1}. The extrapolated value 
is close to the experimental value.   
For the {\it same configurations} we determined then 
these parameters by fitting the three-point correlators shown 
in Fig. 2. In principle the correlator is a function 
of two points, x and y, which we have chosen to be orthogonal. 
 So one can 
separately check that the form factor has no or neglible 
x dependence and then look for the y dependence. 
The x value has to be chosen large enough, so 
that the correlator is clearly dominated by 
the pion. However is x too large the statistical 
errrors become too large and no extraction of 
the parameters of the form factor is possible. 
Therefore in practice we followed an alternative scheme, which 
covers both situations simultaneously and 
furthermore saves some computing time: we fixed the 
ratio of x and y to $x=2y$. In app. B we also give formulas for
the continuum contribution, though we want to stress that
our parameter fixing is totally independent of this 
parametrization since we are considering distances 
$x\gg 0.5\fm$, where these contributions are highly 
suppressed compared to the pole contribution.

 The pion parameters, 
$m_\pi$ and $\lambda_\pi$, found from 2- and 3-point correlators
 agree within the error 
bars. Established consistency, we have further determined the 
mass in the monopole formfactor, which actually
can be done rather accurately. We have found that this mass
(or the pion size) turns
out to be sensitive to the instanton $size$, see Fig.3.
Remarkably enough, the experimental corridor 
(two horizontally dashed lines) indicate
the preferred instanton size to be $\rho=0.35\fm$,
in approximate agreement with 1/3 fm \cite{ins:es82i} and in very good
agreement with lattice results \cite{CGHN_94} pointing out
$\rho=0.35\fm$
as well.

As can be seen in the figure a further 
improvement of the experimental measured 
formfactor, as it is planned \cite{ins:CEBAF}, 
and increasing the accuracy of the instanton 
calculation could provide a very powerful 
tool actually to measure the instanton size.  
This is a remarkable statement since 
it directly connects a physical observable 
like the formfactor to an intrinsic 
property of the QCD vacuum, the size 
of the instantons.

In addition in Fig.7 we compared our full RILM result 
to the one-instanton approach of the  
pion formfactor of Forkel and Nielsen 
\cite{ins:foni}, but now
plotted in coordinate space and 
without Borel transform. As can be seen, 
this approach is reliable for 
distances not larger than $0.4-0.5\fm$.  
The same qualitative behaviour of the two 
approaches was found 
for the pion correlator itself \cite{ins:privat}.

\section{Correlators in interacting instanton vacuum }

  On general grounds, the pion channel is
a strongly attractive one, and it was expected that 
pion properties 
(including form factors) do not depend 
on the details of the instanton ensemble, such as correlations etc.

 In order to check these expectations  
we repeated the calculations  for the 
IILM. Our results are shown in Fig. 5 and 6 for 2- and 3-point
correlators (analogous to
Fig.1,2). Because of the particular interaction assumed, the ensemble 
has an average 
instanton size of ${\bar\rho}=0.42\fm$ 
\cite{ins:tses95a}, so we are not able to 
compare different sizes for this model. However, by 
comparing the IILM with the RILM for 
this common size of ${\bar\rho}=0.42\fm$, we can find out the role of
the correlations. 
As can be seen  from \qutab{tab2}, we have in fact found that the 
pion coupling constant depends rather 
strongly on the model (to a degree, this may be traced 
back simply to different values of the 
quark condensate), while 
the pion and the monopole mass 
(of the form factor) are less dependent.
Within the error bars of our fits, the  
$m_\pi^*$ and $m_V$ of  \qutab{tab2} seem 
to agree with the values of \qutab{tab1} 
for ${\bar\rho}=0.42\fm$.

The same can be done for the isovector scalar meson, 
the $\delta$, as shown in Fig. 4.  However, as already 
mentioned, in this channel the repulsion is very large 
and the simple random instanton liquid  does no longer 
give a good description of the correlator. On very 
general grounds, the correlator should be positive 
for larger distances, whereas the RILM correlator 
changes sign for $x=2y\simeq 0.6\fm$ and crosses the axis 
again at $x=2y\simeq 1.4\fm$.  
The situation changes dramatically if we consider the  
IILM correlator. This one, as can be seen also in Fig. 4, 
stays positive over the whole range of available distances. 
However for $x=2y\simeq 1.8\fm$ the error bars become 
larger and the correlator increases again, which is
an unphysical effect. Since the mass of the $\delta$ is 
too large to be measured precise enough in our model, we 
do not attempt to determine the mass of the 
possible form factor. 
The figure should serve only 
as a further qualitative 
justification of our IILM model, in which 
the correlation and interaction of the instantons 
provide the right behaviour for the strong repulsive 
channels.

   The last issue we address in this work deal with the old question
   of {\it vector dominance}. We remind the reader that it suggests
   that a $complete$ pion formfactor is given by a rho-meson pole
   alone,
and thus the fitted mass in the formfactor $m_V$ is nothing else but
$m_\rho$. It is certainly approximately true in nature: and therefore
we have tried to
check whether indeed there exist a strong correlation between them
(in the models used).  
Therefore we have determined also the parameters of 
the $\rho$ meson correlator (shown in Fig. 8).   
However we found that: (i)  $m_\rho$  
is only very weakly dependent on the 
instanton size, from $875$ to $915\MeV$ 
in our given interval for the instanton sizes; (ii)
$m_\rho$ is more sensitive to correlations than $m_V$; and (iii)
$m_V$  is 
also significantly (more than $10\%$)
smaller
than the experimental $m_\rho$ \footnote{The latter one is 
in the instanton 
model unfortunately too large, probably 
due to the crudeness of our phenomenlogical 
Ansatz. }.  We  
therefore conclude, that in the instanton-based models
the  mass in the pion 
form factor $m_V$ 
has nothing to do with 
the $\rho$-meson mass, and the fact that they are close numerically is
probably
just a coincidence.

%
\section{Summary and Discussion }

We calculated the pion electromagnetic formfactor 
for space like 
momentum transfer, based on a three-point Euclidean correlation function
of two pseudoscalar, isovector currents and an external 
electromagnetic current. Two variants of the 
instanton-based  QCD vacuum models were used, a random 
one, RILM, and the interacting one, IILM. Our main result is the existence
of the direct connection between the size of the instanton and the
size of the pion.

In contrast to earlier work \cite{ins:foni}, we have calculated the
correlation 
function in coordinate space at sufficiently large distances,
clearly separating the pion pole contribution from those of
non-resonance states. We have accurately determined 
 the pion coupling constant $\lambda_\pi$, the pion mass $m_\pi$ and
the parameter $m_V$ of the monopole fit to the electromagnetic formfactor.
The first two parameters are fitted $both$ from 2-point and 3-point
correlators, and the results are consistent within errors.

We have calculated the 
monopole mass of the
formfactor $m_V$
for three different sets
of average instanton sizes at {\it fixed} instanton density 
${\bar n}=1\fm^{-4}$ and find that it is indeed sensitive. Thus the
pion size is directly related to the instanton size. Furthermore,  
the experimental monopole mass 
$m_V\simeq 679\pm 18\MeV$  
is only
reproduced for a mean instanton size around
$0.35\fm$, a value consistent with direct lattice measurements \cite{CGHN_94}.

As a further check we used the streamline version
\cite{ins:stream,ins:tses95a}  of 
 the interacting instanton model (IILM). 
In this case the inverse
monopole mass is small $0.16\fm$, or $m_V\simeq 1200\MeV$.
It does not agree with data but  agrees well with
the RILM result for the same
 mean instanton size of ${\bar\rho}=0.42\fm$. 
These results indicate that  the formfactor does not
 depend very much on the correlations between
instantons. Other quantities like the condensates, e.g., 
do however depend rather strongly on which model is used and therefore 
on such
correlations. So, the IILM can and should be improved. 

We have considered the issue of vector dominance, suggesting that the
formfactor mass $m_V$ is nothing but $m_\rho$. 
However, in our model used,
we have found that both quantities show different dependence on the
instanton size and correlations in the ensemble. We therefore suggest that
vector dominance should not hold in general and seems to be more 
a kind of a coincidence. This will be checked by considering 
other 3-point functions in forthcoming publications.

\acknowledgements
The authors acknowledge the support of Alexander von Humboldt
Foundation (AB) and Department of Energy grant
DE-FG02-88ER40388. Useful discussions with Th. Schaefer (INT)
are acknowledged.

\pagebreak

\begin{table}[h]
\baselineskip 12pt
\begin{center}
\begin{tabular}{c|c|c|c|c|c|c|c|c}
            & \multicolumn{2}{c}{2-point}
            & \multicolumn{2}{c}{3-point}
            &         &    \\
             &  $\lambda_\pi^{1/2}[\MeV]$  &  $m_\pi^*$[\MeV]
             &  $\lambda_\pi^{1/2}[\MeV]$  &  $m_\pi^*$[\MeV]
            &  $m_V$[\MeV]  &
            $ \langle{\bar u}u\rangle^{1/3}[\MeV]$    &
             $m_\pi$ [\MeV]          &
              $f_\pi$[\MeV]          \\
  \hline
$\rho=0.28$ & 455$\pm$5 & 265$\pm$5& 455$\pm$ 5 & 265$\pm$5 & 1250$\pm$50&
    280.3$\pm$ 0.46 &
                          139     &   149.9$\pm$5  \\
$\rho=0.35$ & 490$\pm$5 & 285$\pm$5& 480$\pm$ 5 & 290$\pm$5 & 680$\pm$20 &
     259.1$\pm$ 0.25 &
                           150    &    106.6$\pm$5  \\
$\rho=0.42$ & 520$\pm$5 & 255$\pm$5& 510$\pm$ 5 & 265$\pm$5 & 550$\pm$20 &
      245.2$\pm$ 0.39 &
                            136   &    79.9$\pm$5   \\
\end{tabular}
 \end{center}
 \caption{\label{tab1}The pion properties in the RILM
 for different values of the
 instanton size $\rho$. The pion coupling constant $\lambda_\pi$
 and pion mass $m_\pi^*$ for $m_u=0.1\Lambda\simeq 20\MeV$
 obtained from the 2-point and
 3-point function are determined independently and coincide within
 the uncertainty of the fit. For comparison, Forkel and Nielsson
 used $\lambda_\pi^{1/2}=363\MeV$, which
 either corresponds to a rather low value for the condensate
 $\langle{\bar u}u\rangle^{1/3}\simeq 200\MeV$ or
 a large $f_\pi\simeq 180\MeV$.  The physical pion mass $m_\pi$ in the table follows
 from an assumed Gell-Mann Oakes Renner scaling relation with
 $m_u+m_d=11\MeV$.  }
\end{table}
\begin{table}[h]
\baselineskip 12pt
\begin{center}
\begin{tabular}{c|c|c|c|c|c|c|c|c}
            & \multicolumn{2}{c}{2-point}
            & \multicolumn{2}{c}{3-point}
            &         &    \\
             &  $\lambda_\pi^{1/2}[\MeV]$  &  $m_\pi^*$[\MeV]
             &  $\lambda_\pi^{1/2}[\MeV]$  &  $m_\pi^*$[\MeV]
            &  $m_V$[\MeV]  &
            $ \langle{\bar u}u\rangle^{1/3}[\MeV]$    &
             $m_\pi$ [\MeV]          &
              $f_\pi$[\MeV]          \\
  \hline
${\bar\rho}=0.42 $ & 395$\pm$5 & 260$\pm$5& 370$\pm$ 5 & 295$\pm$5 & 1200$\pm$50&
     217.3 $\pm$ 0.55  &
                          113     &  93  $\pm$5  \\
\end{tabular}
\end{center}
 \caption{\label{tab2}The pion properties in the IILM
 for $\Lambda=306\MeV$. The pion coupling constant $\lambda_\pi$
 and pion mass $m_\pi^*$ for $m_u=0.1\Lambda$
 obtained from the 2-point and
 3-point function are determined independently.
 The physical pion mass $m_\pi$ in the table follows
 from an assumed Gell-Mann Oakes Renner scaling relation with
 $m_u+m_d=11\MeV$.  }
\end{table}

\appendix


\section{One instanton formulas }

\subsection{The pion correlator }

The pion correlator for pseudoscalar currents $j_5(x)$
for short distances in the asymptotic regime  is simply given by
\beq   <0\mid T j_5(x) j_5(0) \mid 0>
    = {N_c \over  \pi^4} {1\over x^6}  := \Pi(x)_{free}
\label{b1}
\eeq
Corrections to this correlator at larger distances are given by
the direct instanton contribution:
\bea     <0\mid T j_5(x) j_5(0) \mid 0>
       &=&  {12\over \pi^2}
      {\bar n} {1\over {\bar m}^2} \rho^4
      {x^2\over \sqrt{4x^2\rho^2+x^4}^7 }
        \nn  
     \left[ {\rm arctanh} {x^2\over \sqrt{4x^2\rho^2+x^4}}
       8 \left( x^4+2\rho^2+2\rho^2x^2 \right)
           \right.  
   & &       \left.
    + \sqrt{4x^2\rho^2+x^4}  {1\over 3\rho^2}
        \left( 2x^4-10x^2\rho^2-12\rho^4\right)
         \right]  \nn
     &:=&  \Pi(x)_{inst}
\label{b2}
\eea
where ${\bar m}$ (cf. eq. \ref{mbardef}) is the effective constituent 
quark mass \cite{SVZ_80,CDG_78}
The asymptotics of the one-instanton formula \qeq{b2} for short distances
is given by
\beq   \lim_{x\to 0} \Pi(x)_{inst}
        =         { 1 \over \pi^2}
         { {\bar n } \over {\bar m}^2} \rho^4
          {2 \over 5  \rho^8}
\label{b3}
\eeq
which is suppressed by the free contribution \qeq{b1} with
the power $x^6$, so that the sum of   $\Pi(x)_{inst}$
and $\Pi(x)_{free}$, normalized with
$\Pi(x)_{free}$, goes to unity. This is also the case for
our instanton liquid ensemble of  Fig.1 and 5.  
For large distances the limit of \qeq{b2}  becomes
\beq
      \lim_{x\to\infty} \Pi(x)_{inst}
        =         {12\over \pi^2}
         {\bar n} {1\over {\bar m}^2} \rho^4
          {2\over 3 \rho^2 x^6 }
\label{b4}
\eeq
Here the  zero mode contribution is proportional to the free part
\qeq{b1}
and actually dominates the sum of both expression.
For completeness the pole contribution of the 2-point function 
in a simple pole plus continuum Ansatz  
is \quref{es93b} given by
\beq    \Pi(x)_{pole} = { \lambda_\pi^2 \over 4 \pi^2 }
        { 1 \over x^2 } m_\pi \mid x\mid K_1(m_\pi\mid x\mid)
\label{b5}
\eeq
whereas the continuum results \quref{es93b} reads 
\beq  \Pi(x)_{cont} = {3\over 16 \pi^6 } \left(
      K_1(E x) (16E x + 4 (Ex)^3)
     + K_0(E x) ( 8 (E x)^2 + ( E x)^4 ) \right)
\label{b6}
\eeq
Indeed the small size limit of \qeq{b6} for a fixed threshold $E$ is 
given by
\beq   \lim_{Ex\to 0} \Pi(x)_{cont} =  \Pi(x)_{free} \eeq
whereas the normalized pole contribution, i.e.  
$\lim_{x\to 0}\Pi(x)_{pole}/\Pi(x)_{free}$, vanishes.

\subsection{The pion EM-formfactor}

From our general expression \qeq{g21} and  
the one-instanton Ansatz \qeq{zeroeff} the correlator is given by
\bea
      & &   \Pi_\mu (x,y) =
         {- i e_\pi \over  \pi^6 } { 4 {\bar n} \rho^4 \over {\bar m}^2 }
             { 1 \over  (y^2 + x^2/4)^2 }
             \int d^4 z   \nn & &
            { y_\mu (x^2/4+2yz-z^2)
          + x_\mu/2 ( -y^2 + 2yz-z^2)
          + z_\mu ( -y^2 - x^2/4) \over
         \mid y-z\mid ((y-z)^2+\rho^2)^{1.5}
          ((x/2-z)^2+\rho^2)^{3} \mid x/2+z\mid ((x/2+z)^2+\rho^2)^{1.5}
          }
\label{c1}
\eea 
where the trivial averaging over color orientations is already done.
The averaging over the instanton position in \qeq{c1}  can be
performed after introducing 5 Feynman parameter integrals
for the 5 denominators in \qeq{c1}. After this, it is even
possible to perform 2 of the Feynman integrals analytically,
which results in the rather awkward expression
\bea
       &  &  \Pi_\mu (x,y)     =
         {e_\pi  \over \pi^8 } { {\bar n} \rho^4 \over {\bar m}^2 }
         5760
         { - i x_\mu  \over (y^2 + x^2/4)^2  }
         \int_0^1 d\ a_1 \int_0^{1-a1} d\ a_2 \int_0^{1-a1-a2} d a_3
          a_1^2
         \sqrt{ {a_2 \over a_3} }  \nn
     & &  \left[    {  \pi^3 (1-a_1-a_2)  \over 3840 }
          { 35 b^3 +120 a b^2 + 144 b a^2 + 64 a^3 \over
              a^{4.5} (a+b)^{3.5}   }  \right.    \nn
            & &
        \left( x^2/4  ( (a_1-a_2-a_3)^2 + (a_1-a_2-a_3)) + y^2
              ( (a_1+a_2+a_3)^2 + (a1-a_2-a_3) )       \right)
                \nn      &   &   \left.
           + { \pi^3  (1-a_1-a_2) \over 960 }
       { 5 b^2 + 12 b a + 8 a^2 \over a^{3.5} (a+b)^{2.5} }
         \right]
\label{c2}
\eea
where
\beq   a=\rho^2(1-a_3) + x^2/4  ((a_1+a_2+a_3) - (a_1-a_2-a_3)^2 )
          + y^2 ((a_1+a_2+a_3) - (a_1+a_2+a_3)^2 )
\eeq
and $b=-\rho^2(1-a_1-a_2-a_3)$.
Similiar to the 2-point function one can consider the
short distance behaviour of this function and finds
\beq
           \lim_{x,y\to 0}
          \Gamma_\mu (x,y,0) = -
       { e_\pi \over 5 \pi^4 } { {\bar n}  \over m^2 \rho^4 }
                    { i x_\mu \over (y^2 + x^2/4)^2  }
\label{c3}
\eeq
Because our triangular diagram for the EM form factor
in this approximation actually consists of
2 zero mode propagators and one free propagator, the
zero mode part reduces to a constant,
similiar to \qeq{b3}, and a single quark propagator
in \qeq{c3} remains.
The full behaviour of \qeq{c2} can be seen in Fig. 7
compared to the full (RILM) result. Obviously 
the one instanton formula deviates
from the full result for distances larger than $0.4\fm$.


\section{Continuum contribution}

For the continuum contribution one has to calculate the
double discontinuity of the spectral density
from the expression for the triangular diagram
according to the Cutkovsky rule as
 \beq  \rho_\mu (s_1,s_2,Q^2) =  {N_c \over \pi^2}{1\over 4 \pi}
         \int {d^4 k\over (2\pi)^4 }
        \theta(k_0) \delta (k^2)
         \theta(k_0-p_0) \delta ((k-p)^2)
         \theta(k_0-p_0') \delta ((k-p')^2)
\eeq
with the result \quref{iosm}
\beq
        \rho_\mu (s_1,s_2,-q^2) = { s_1 s_2  \over  \pi^2 }
         { N_c  \over \lambda^{1.5}      }
           \left[
           {\bar  p}_\mu   q^2  -  q_\mu(s_2 - s_1)/4
           \right]
\eeq
where
$\lambda=(s_1+s_2+q^2)^2 - 4s_1 s_2$.
Inserting this into the dispersion relation and
transforming the resulting expression
into Euclidean space,  one can use
the Feynman parameters as before for the pole contribution
to
perform the ${\bar p}$-integration analytically.
In order to model the continuum one usually introduces
a threshold parameter $s_0$ into dispersion relations. If one 
does it,
the continuum contribution
reduces to a four
dimensional integral, which can be easily done
numerically.
It reads
\bea     & &  \Pi_{\mu,cont} (x,y)
           = -i N_c
        \int_0^\infty  ds_1 
         ds_2    \int_0^1  d\alpha  \int_0^\infty  dq
        \ \left(1  -  \theta( s_1 - s_0) \theta( s_2 - s_0) \right)
           {N_c \over 64 \pi^6 }
        { s_1 s_2 \  q^2   \over \lambda^{1.5} }
          \nn  & &
        \left[
        q^2 M  {  x_\mu  \over \mid x\mid } K_1(\mid x\mid M)
      {   J_1( q \mid y + x(0.5-\alpha)  \mid )
            \over
            \mid y + x (0.5-\alpha)\mid
                    }
         \right.  \nn
          & &    \left.
         {   (s'-s)  \over 4 }    K_0(\mid x\mid M)
        { 1 \over  \mid y + x (0.5-\alpha)\mid^2  }
        \left(
          J_0           ( q  \mid y+x(0.5-\alpha)\mid )
         q (  y_\mu
              + x_\mu  (0.5-\alpha)  )
        \right. \right.  \nn & & \left. \left.
        - 2  J_1 (q \mid y+x(0.5-\alpha)\mid )
          {     y_\mu
              + x_\mu (0.5-\alpha)
                \over \mid y + x(0.5 - \alpha) \mid   }
         \right)
           \right]
    \label{a1}
\eea
where
\beq
       M^2 = q^2 \alpha(1-\alpha) + s \alpha + (1-\alpha) s'  >  0
\eeq
and $\lambda=(s_1+s_2+q^2)^2 - 4 s_1 s_2$.
In \qeq{a1} we introduced a simple model for the continuum
by introducing the $\theta$-functions \quref{iosm2}. Below we
argue why their detailed form is not important here.

Now considering our
preferred geometrical arrangement $x\ y=0$ it is clear from
\qeq{a1} that the $x_\mu$ part in the last two lines
vanishes due to the $\alpha$-integration.  Finally the
$y_\mu$ terms vanishes because these
are antisymmetric in $s\leftrightarrow s'$ \footnote{Using that
$M^2$ is invariant under simultaneous change of
$s\leftrightarrow s'$    and
$\alpha\leftrightarrow(1-\alpha)$ as well as the remainder of the
formula \qeq{a1}.}.
Therefore only the
first term in \qeq{a1} finally contributes, so that
\bea  & &   \Pi_{\mu,pole}(x,y) +   \Pi_{\mu,cont}(x,y)  =
        \nn & &
       -i \int_0^\infty dq   \int_0^1 d\alpha
       { q^2 \over 32 \pi^4 }
%
       \left[
       { J_1 ( q \mid y-x(0.5-\alpha)\mid )
       \over \mid y - x(0.5 - \alpha) \mid   }
        \sqrt{m_\pi^2 +q^2\alpha(1-\alpha)}  \right.
         \nn & & \left.
       K_1
       \left( \mid x \mid \sqrt{m_\pi^2 +q^2\alpha(1-\alpha) } \right)
       {    2 \lambda_\pi^2  e_\pi
           \over 1 + q^2/m_V^2 }
         {x_\mu \over \mid x \mid }
         \right.      \nn  &   &       \left.
         +   \int_0^\infty  ds_1    \int_0^\infty  ds_2
         \left(1 -  \theta( s_1 - s_0) \theta( s_2 - s_0) \right)
         {N_c\over 2 \pi^2}
        { s_1 s_2 \ q^2  \over \lambda^{1.5} }
         {  x_\mu M \over  \mid x\mid }
          \right.    \nn  & &  \left.      
         K_1\left(\mid x\mid M\right)
       {  J_1\left( q \mid y + x(0.5-\alpha)  \mid \right)  \over
        \mid y - x(0.5 - \alpha) \mid  }
       \right]
      \label{a4}
 \eea
Similiar to the 2-point function \qeq{b6} the small size limit
of \qeq{a4}, which is actually the small size limit of
$\Pi_{\mu,cont}(x,y)$, is given by  $\Pi_{\mu}(x,y)$ of \qeq{freeprop}.
First one should note that \qeq{a4} is finite with
neither referring to subtraction terms, which would be necessary in
momentum space due to the insufficient fast decreasing spectral
density, nor to Borel transformations. The latter method
eliminates the subtraction terms, because these are
known to be polynomials in either $p$ or $p'$ and therefore
vanish after taking derivatives finite times \quref{iosm2}.
In the present approach, the fourier transform of the
dispersion relation into coordinate space, which is finite
\'a priori, circumvents the introduction of any subtractions.
However one should also stress, that the continuum contribution
for a threshold of $s_0>1\GeV$
is highly suppressed for distances above $1\fm$. We have given those 
expressions for completeness and their inclusion does not
affect our parameter fixing at all.

\pagebreak


\pagebreak
\noindent
\hskip 1.9in
\epsfig{file=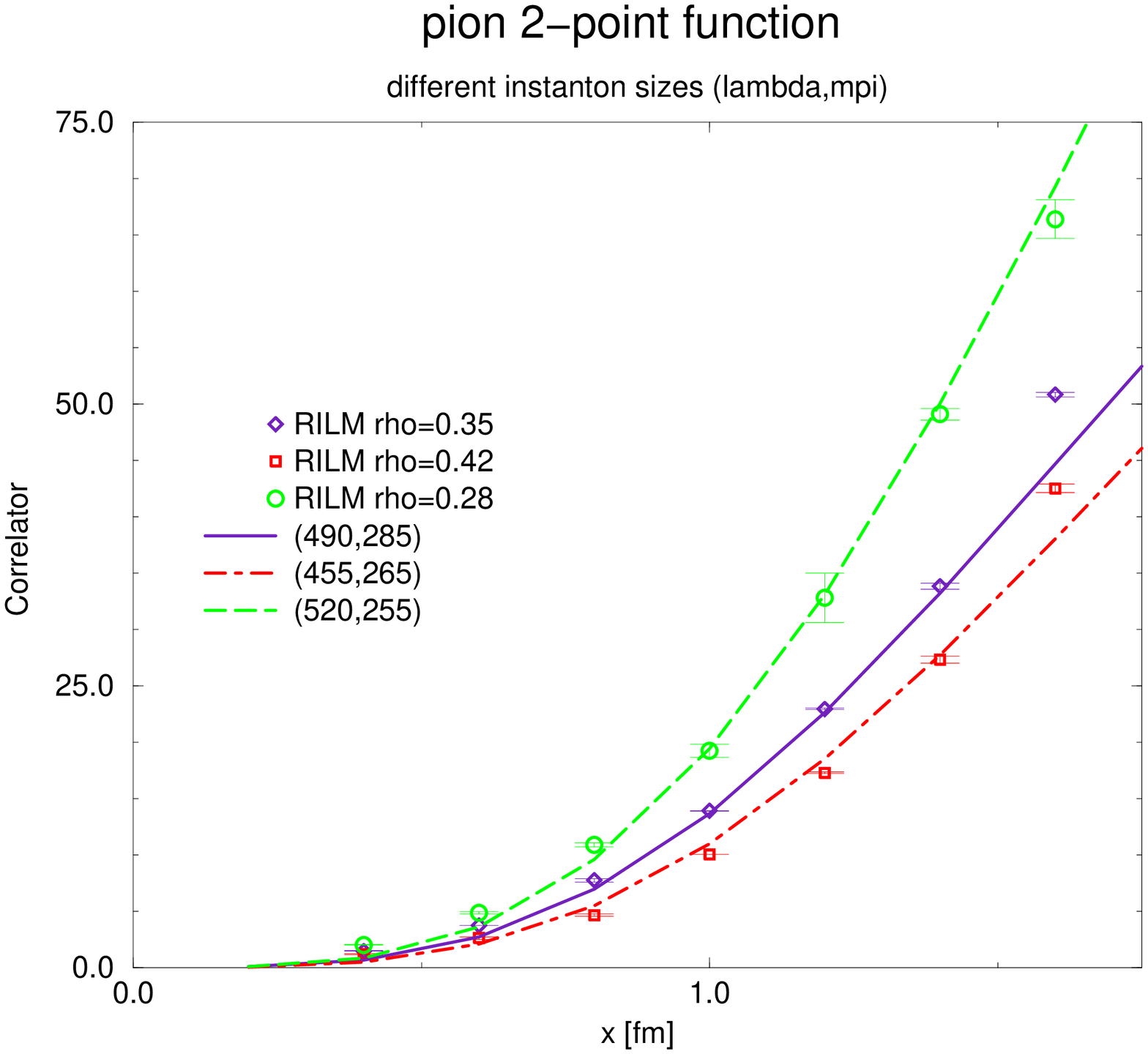,width=12cm,angle=-90}
\begin{center}
\parbox{6in}{
\small Fig.~1:\quad The two-point correlation function of the
pion, normalized to the free correlator and for three
different instanton sizes in the random configuration (RILM). The
instanton
density is kept constant ${\bar n}=1\fm^{-4}$. The parameters of the
pole contribution, $\lambda_\pi$
and $m_\pi$, are given in the legend.}
\end{center}

\pagebreak 
\hskip 1.9in
 \epsfig{file=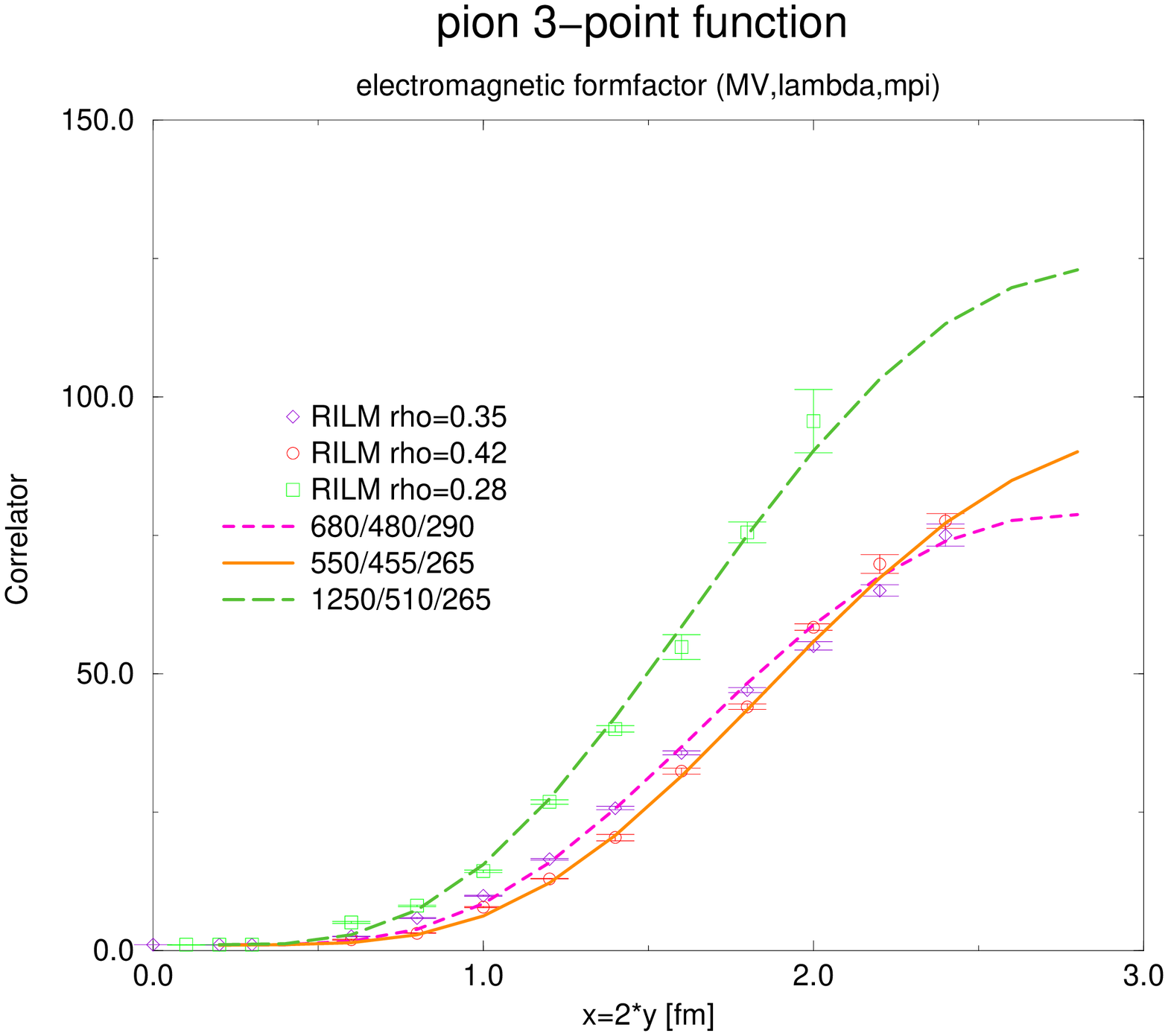,width=12cm,angle=-90}
\begin{center}
\parbox{6in}{
\small Fig.~2:\quad The three-point correlation function of the
pion and an external electromagnetic current, normalized to the free
correlator and for
three different instanton sizes in the random configuration (RILM). The
instanton
density is kept constant ${\bar n}=1\fm^{-4}$. The parameters of the
pole contribution, $m_V$, $\lambda_\pi$
and $m_\pi$ in this order, are given in the legend.}
\end{center}
\pagebreak
%
\hskip 1.9in
 \epsfig{file=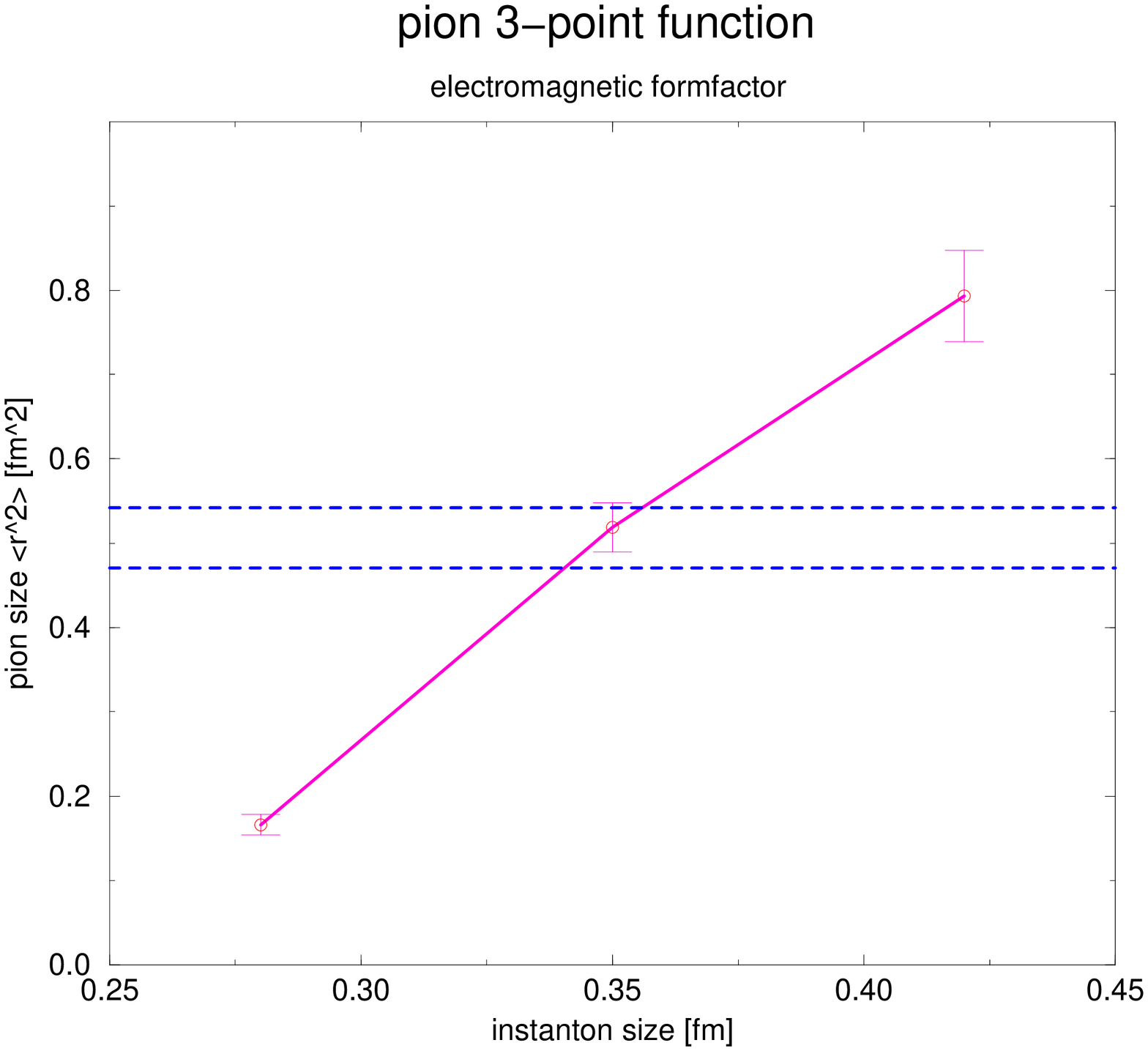,width=12cm,angle=-90}
\begin{center}
\parbox{6in}{
\small Fig.~3:\quad The squared electromagnetic radius 
of the  pion as a function of
instanton size for the random ensemble (RILM).
The instanton
density is kept constant ${\bar n}=1\fm^{-4}$. }
\end{center}
\pagebreak
\hskip 1.9in
 \epsfig{file=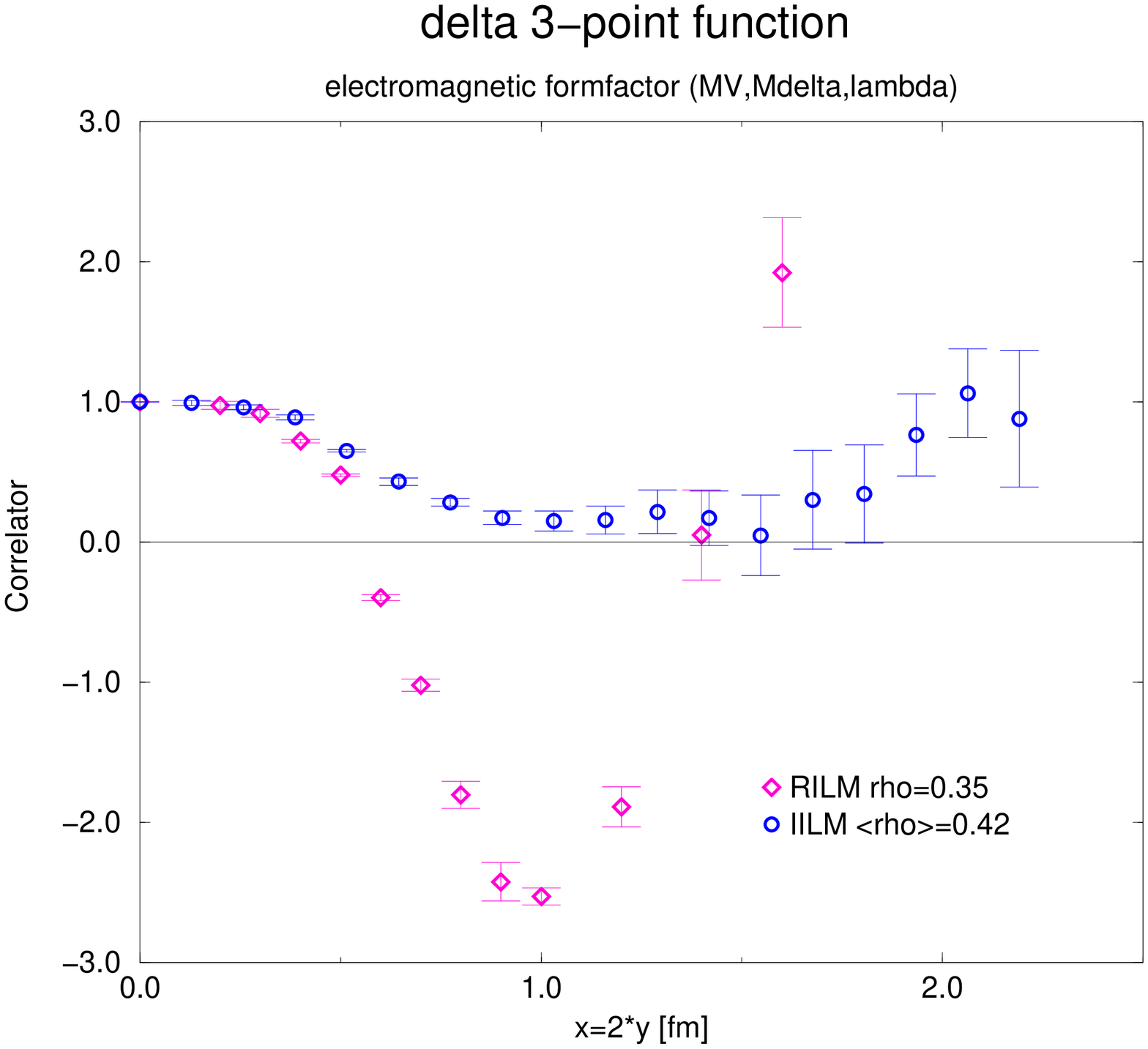,width=12cm,angle=-90}
\begin{center}
\parbox{6in}{
\small Fig.~4:\quad
The three-point correlation function of the
delta (scalar, isovector) and an external electromagnetic current,
normalized to the free correlator
in the random configuration (RILM)
and the interacting ensemble (IILM).
The instanton
density is kept constant ${\bar n}=1\fm^{-4}$. }
\end{center}
\pagebreak
%
\hskip 1.9in
 \epsfig{file=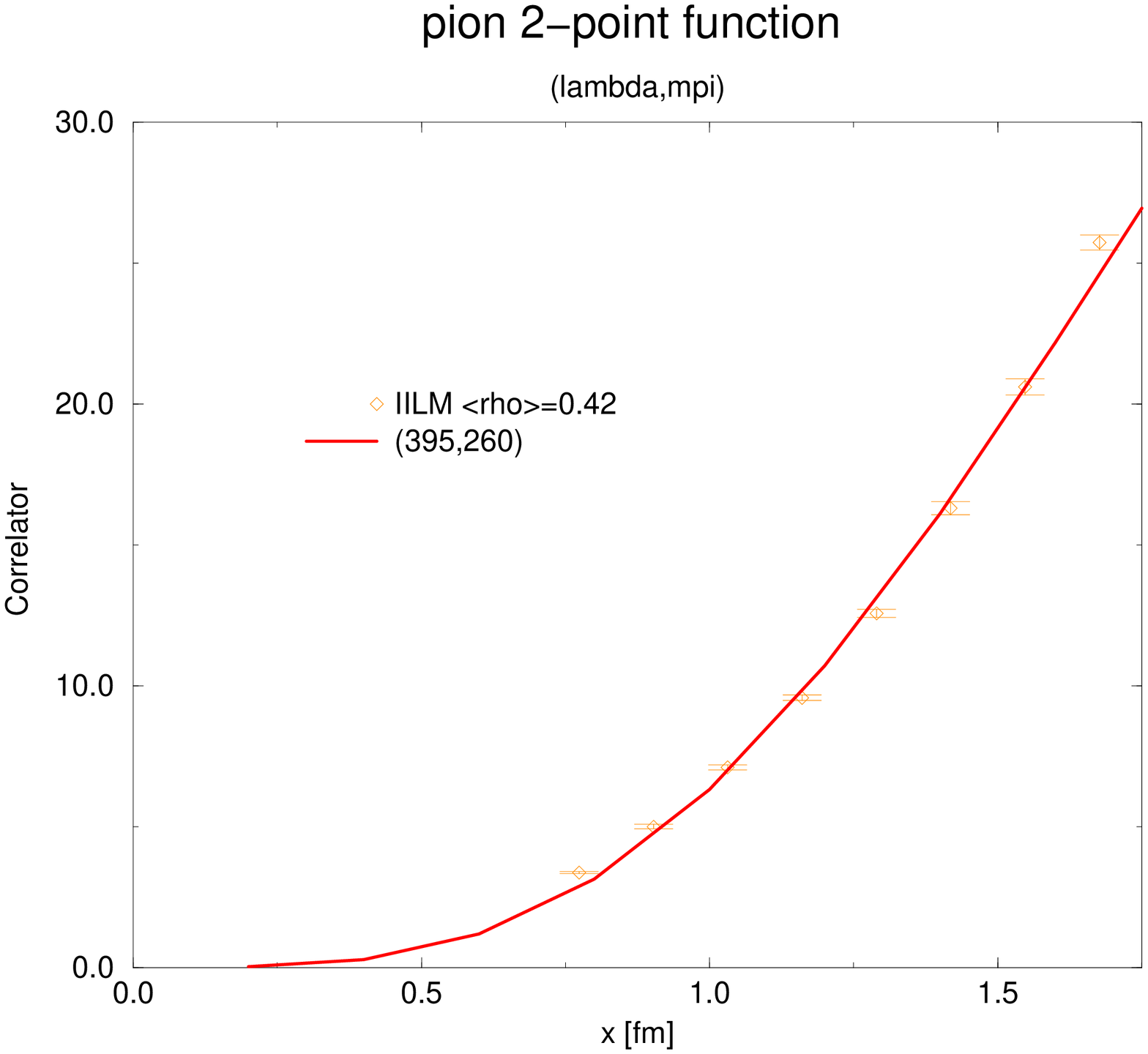,width=12cm,angle=-90}
\begin{center}
\parbox{6in}{
\small Fig.~5:\quad The two-point correlation function of the
pion, normalized to the free correlator and
in the interacting configuration (IILM). The average
instanton size is ${\bar\rho}=0.42\fm$ and
instanton density is $\rho=1\fm^{-4}$.}
\end{center}
\pagebreak
\hskip 1.9in
 \epsfig{file=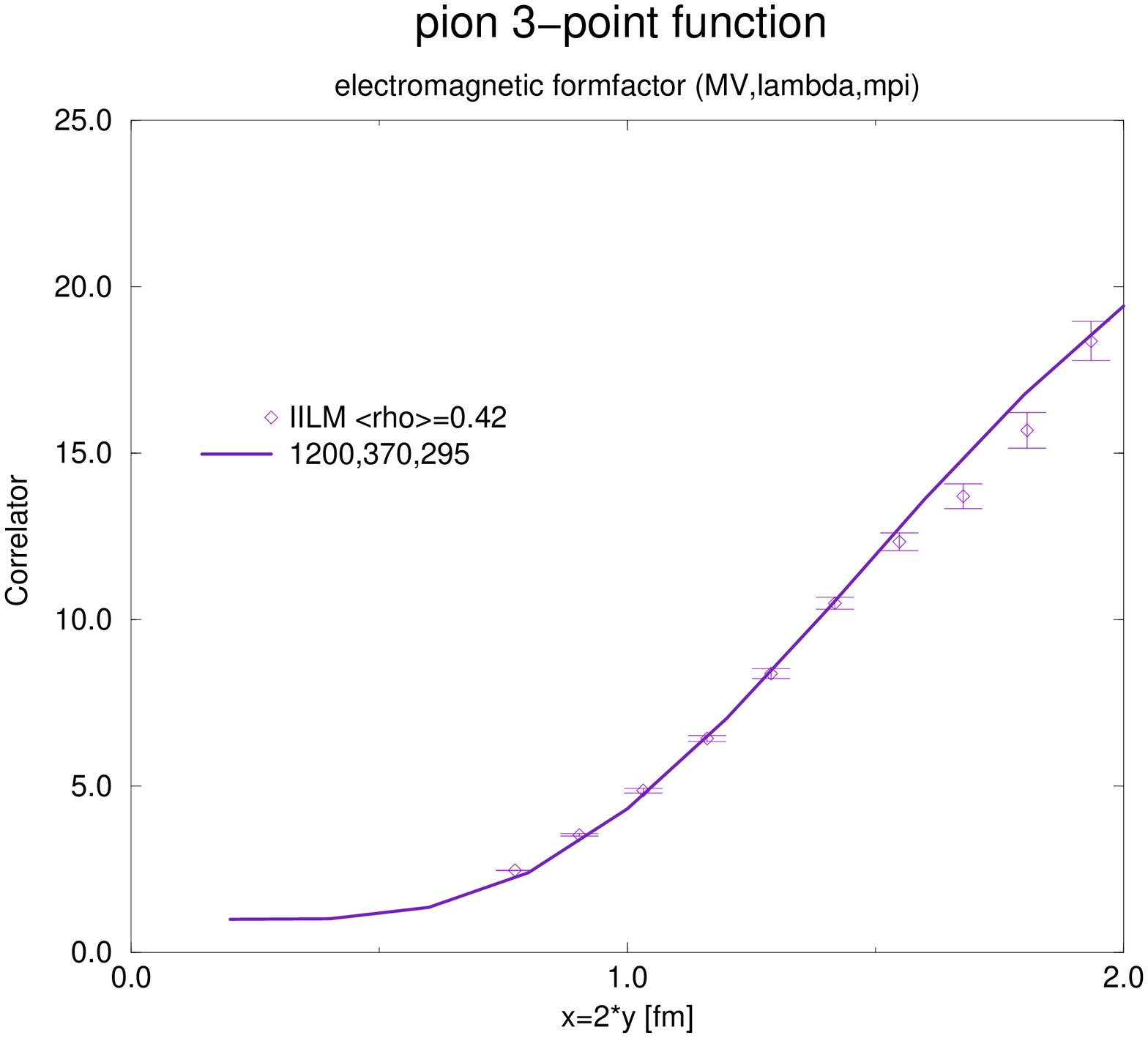,width=12cm,angle=-90}
\begin{center}
\parbox{6in}{
\small Fig.~6:\quad The three-point correlation function of the
pion and an external electromagnetic current, normalized to the free
correlator and
for the interacting configuration (IILM).
The averaged
instanton size is ${\bar\rho}=0.42\fm$ and
the instanton density is $\rho=1\fm^{-4}$.  }
\end{center}
%
\pagebreak
\hskip 1.9in
 \epsfig{file=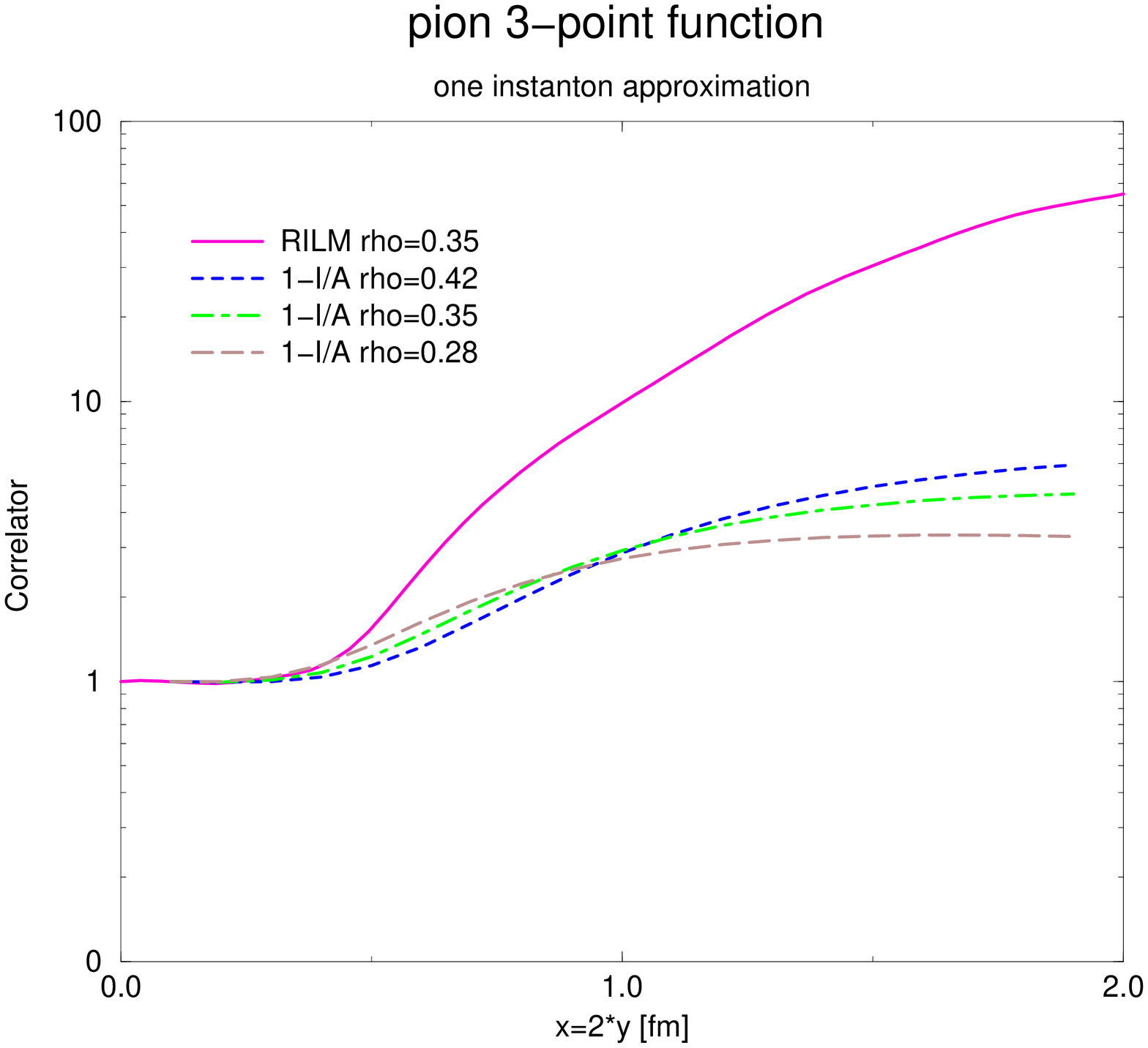,width=12cm,angle=-90}
\begin{center}
\parbox{6in}{
\small Fig.~7:\quad  The three-point correlation function of the
pion and an external electromagnetic current
for the RILM and the one-instanton (1inst) formula for
three different instanton sizes. }
\end{center}
\pagebreak
\hskip 1.9in
 \epsfig{file=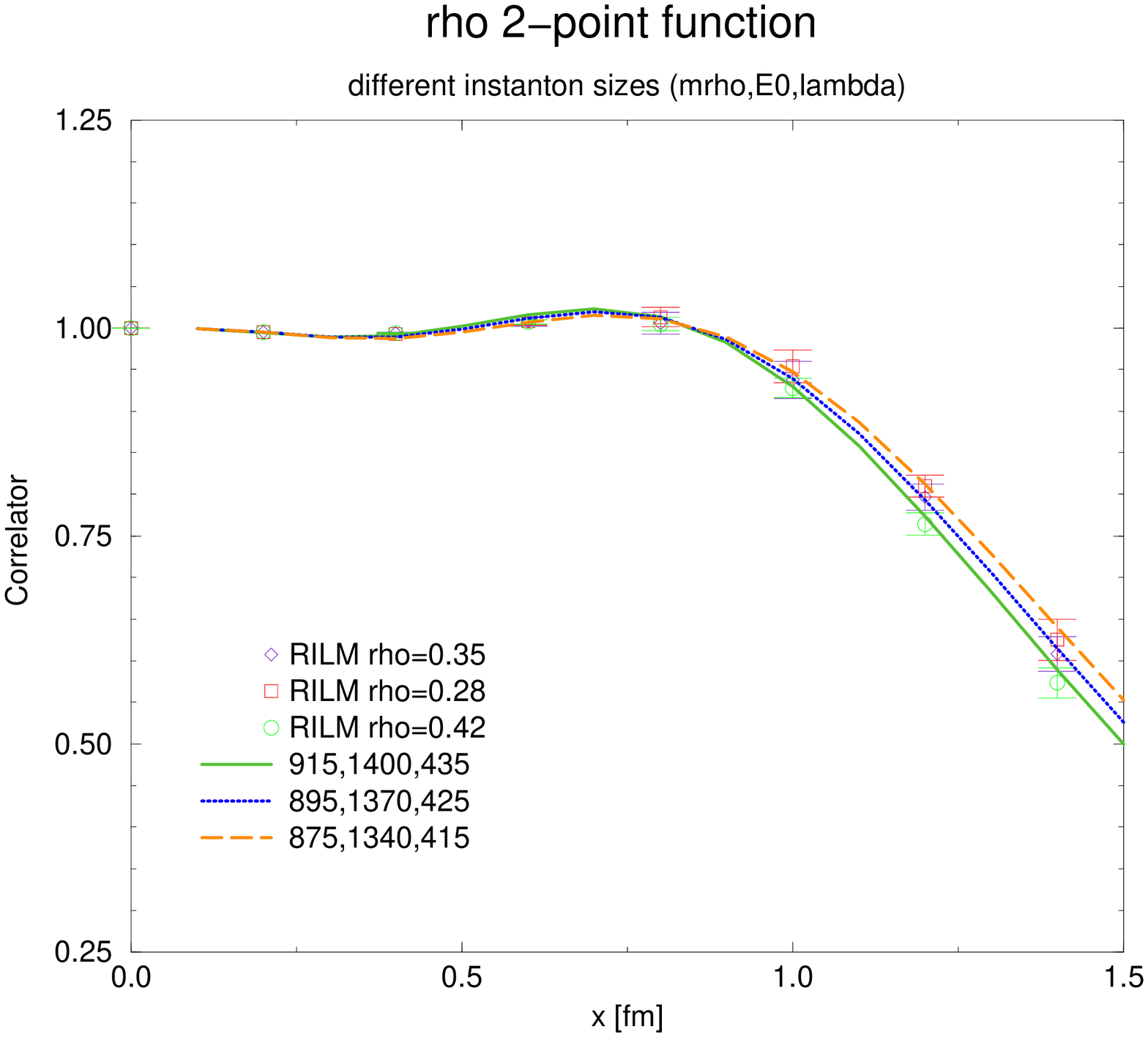,width=12cm,angle=-90}
\begin{center}
\parbox{6in}{
\small Fig.~8:\quad The two-point correlation function of the
$\rho$-meson
for the RILM and three different instanton sizes, normalized to
the two-point correlation function. In the legend, the 
$\rho$ meson mass, the threshold parameter $E_0$ and the $\rho$ 
coupling constant are given.  
     }
\end{center}

\end{document}